\documentclass[10pt]{article}
\usepackage{amssymb}
\usepackage{graphicx}
\usepackage{subfigure}
\usepackage{amsmath}
\usepackage{cite}
\usepackage{color}

\date{}

\begin{document}

\begin{flushleft}
{\Large
\textbf{How Random are Online Social Interactions?}
}
\\
Chunyan Wang$^{1}$ and
Bernardo A. Huberman$^{2,\ast}$
\\
\bf{1} Department of Applied Physics, Stanford University, California, USA
\\
\bf{2} Social Computing Group, HP Labs, California, USA
\\
$\ast$ E-mail: bernardo.huberman@hp.com
\end{flushleft}

\section*{Abstract}

The massive amounts of data that social media generates has facilitated the study of online human behavior on a scale unimaginable a few years ago. At the same time, the much discussed apparent randomness with which people interact online makes it appear as if these studies cannot reveal predictive social behaviors that could be used for developing better platforms and services.  We use two large social databases to measure the mutual information entropy that both individual and group actions generate as they evolve over time. We show that user's interaction sequences have strong deterministic components, in contrast with existing assumptions and models. In addition, we show that individual interactions are more predictable when users act on their own rather than when attending  group activities.

\section*{Introduction}

Recent developments in digital technology have made possible the collection and analysis of massive amount of human social data and the ensuing discovery of a number of strong online behavioral patterns~\cite{webB01,humanV06,webB06,webF07,humanM08,webR10,humanW10,webS11,webH98,webT03,webG11}. These patterns are important for two reasons. First, they yield predictions about future behavior that can be incorporated into the design of useful social media and services, and second, they provide an empirical test of the many social  theoretical models that have been proposed in the literature. As an example, the assumption that events in web traffic data are described by a series of Poisson process ~\cite{humanS03} was shown to be contradicted by measurements of the the waiting time between two consecutive events, which display power law scaling. These power laws are ubiquitous and appear in the analysis of  email exchanges~\cite{burstyE04,burstyB05,burstyR09} and web browsing~\cite{burstyC08,burstyE09,burstyA10}. On the other hand, regular behavioral patterns in real life are a well known phenomenon, as exemplified by vehicular traffic patterns, daily routines, work schedules and the seasonality of economic transactions. At the aggregate level, these regularities are often induced by spatial and temporal constraints, such as the disposition of roads and streets in urban settings or the timing of  daily routines. Other examples are provided by the existence of deterministic patterns in human daily communication~\cite{burstyE04,convpredT11} and phone call location sequences~\cite{mobpredS10}.

When it comes to human online activities many theoretical studies curiously assume uncorrelated random events on the part of the users~\cite{rwP98,humanS03,rwB08,rwA10} which makes their behavior rather unpredictable. Moreover, that literature assumes that a user's future partners in comments and reviews, or how web pages are visited are independent of the history of the process or at best on the previous time step. While these assumptions work well for page ranking in web searching~\cite{rwP98}, online recommendation systems~\cite{rwB08}, link prediction~\cite{linkL11}, and advertising~\cite{rwA10}, it is not clear that they apply to more interactive processes such as contacting friends within online social networks, participating in online discourse and exchanges of email and text messages. Even in cases where a Markovian assumption seems to yield good results, the discovery of deterministic components to online browsing and searching can improve existing algorithms~\cite{webmarkovF12}.

In this paper we study the predictability of online interactions both at the group and individual levels.  To this end, we measure the predictability of online user behavior by using information-theoretic methods applied to real time data of online user activities. This is in the same spirit as a recent study of offline conversations within an organization~\cite{convpredT11}. Using ideas first articulated in studies of gene expressions~\cite{mutualinfoS02},  predictability is here defined as the degree to which one can forecast a  user's interacions based on observations of his previous activity. The main focus of this study is to be contrasted to existing studies of online social behavior, such as recommender systems~\cite{rwB08} and link prediction~\cite{linkL11}, which use statistical learning models to improve the prediction accuracy of novel links and recommendations.  By examining datasets from user commenting activities and place visiting logs, we found that the observed activity sequences deviate from a random walk model with deterministic components. Furthermore, we also compared the predictability of activities when individuals act alone as opposed to as members of a group. In contrast to many model assumptions in studies of online communites and group behavior~\cite{groupW07,groupP07,groupD09}, we observed that individuals are less predictable when attending group or social activities than when acting on their own.

\section*{Methods and Data}
We examined the predictability of online user behavior using datasets from two different websites: Epinions and Whrrl. Epinions is a who-trust-who consumer review site, where users write their personal reviews of a wide variety of products, ranging from automobiles to media (including music, books and movies).  Each user can comment on other users' reviews or comments. The thread of comments forms a conversation of two or more users. To trace the predictability of commenting partners, we collected $88,859$ unique users' comments from the website. For each user, we used the website's API to collect all of their comments with a time stamp for each comment. In total, we gathered $286,317$ threads of comments from different categories containing $722,475$ user comments. The other dataset that we used is from Whrrl.com. Whrrl is a popular LBSN (Location Based Social Network) that people use to explore, rate and share points-of-interest. It also allows users to declare friendships with each other and to interact through visits and check-ins at physical places. Users can check in by using a mobile application on a GPS-equipped smart phone. The types of places that are often visited include restaurants, hotels and bookstores. 

A distinctive feature of this dataset is that a user can check-in by herself or with a group of other people, thus providing a forum for social activities. Users of the site are identified by unique user-ids. In our study, we crawled a friendship network consisting of $24,002$ users and $145,228$ social ties and collected the check-in records of these users' activities from January 2009 to January 2011. The resulting undirected graph had an average degree of $12.101$ and an average shortest-path length of $4.718$, which is typical of a small-world social network. In our observational period of 2 years, there were $357,393$ check-in records over $120,726$ different places associated with these users. For each check-in record, we also collected information such as the exact location (i.e., longitude and latitude), time of check-in and the users involved (i.e., there may be more than one user-id for group check-ins). We were thus able to obtain a series of places the users visited in chronological order.

\begin{figure*}[htl]
 \centering
 \includegraphics[width=8cm]{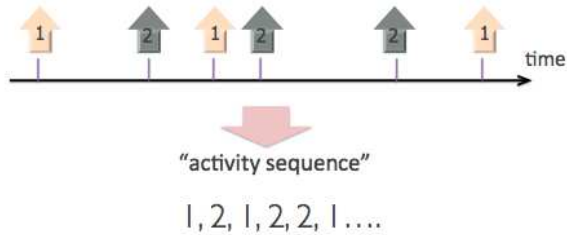}
 \caption{Online activity sequence of a sample user.  Every short vertical line in the figure represents the time of a user activity. There are two observed states for the sampled user's activity sequence, state $1$ and state $2$. The second order correlation, or predictability, of this sequence is measured through the conditional entropy.}
 \label{fig:activity_sequence_sample}
\end{figure*}

The activity sequence is obtained by neglecting the absolute timing of events in the raw dataset. To generate the activity sequence of a certain user, we first sifted out all the events that are associated with the user and we then listed the chronologically ordered sequence of states identified by a unique number. For the Whrrl dataset, we labeled each activity as a group one if the user was checking in with others. To determine the extent to which user behavior is predictable we used standard information-theoretic methods similar to those used in the analysis of gene expression~\cite{mutualinfoS02,biascorrS07}. For instance, we consider a user $A$ as having $M_A$ possible states, where each state in the sequence can correspond to either an online conversation partner or a check-in location. An example of a user's activity sequence is shown in Figure~\ref{fig:activity_sequence_sample}, where two states, $1$ and $2$, form the sequence. We then used the observed sequences to examine the degree of second order dependences, which signal the extend to which activities depart from random interactions.

We used entropy to measure the randomness of a user $A$'s activities. The estimated probabilities for all states $p_A(i)$ have the property that  $\sum\limits_{i = 1}^{{M_A}} {{p_A}(i) = 1} $. If we assume that these probabilities do not change with time, the randomness of user $A$'s possible states can be measured by the uncorrelated entropy, defined as 
\begin{equation}
H_A^1 =  - \sum\limits_{i = 1}^{{M_A}} {p_A(i)\log p_A(i)}. 
\end{equation}
Notice that if each state is equally probable, this uncorrelated entropy is maximal and equal to
\begin{equation}
{H_A}^0 = log{M_A}.
\end{equation}
To measure the randomness of the sequence from knowledge of the previous states we introduce the conditional entropy, defined as 
\begin{equation}
H_A^2(i|j) =  - \sum\limits_{j = 1}^{{M_A}} {{p_A}(j)\sum\limits_{j = 1}^{{M_A}} {{p_A}(i|j)} } \log {p_A}(i|j).
\end{equation}
And we quantify the predictability of the user's activity sequence by using the mutual information
\begin{equation}
{I_A} = H_A^1(i) - H_A^2(i|j).
\end{equation}
For each user, the inequalities $0 \le H_A^2 \le H_A^1 \le H_A^0$ are satisfied. $I_A$ is equal to the amount of information one can gain about the next state by knowing the current state. If there is no second order correlation between state sequences, $H_A^1$ is equal to $H_A^2$, and $I_A$ takes the minimum value of $0$. If the next state is completely determined by the previous state, or in other words the user activity is completely predictable,  $I_A$ takes the maximum value of $H_A^1$.

\begin{figure*}[htl]
 \centering
  \subfigure[Conversation] {\includegraphics[width=7cm]{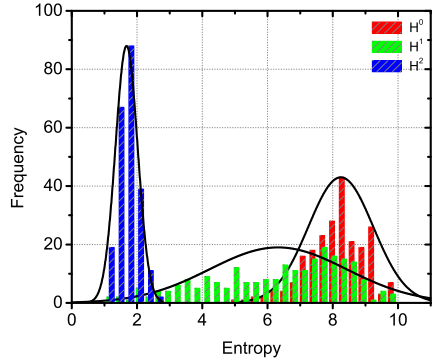}}
  \subfigure[Location Check-in]{\includegraphics[width=7cm]{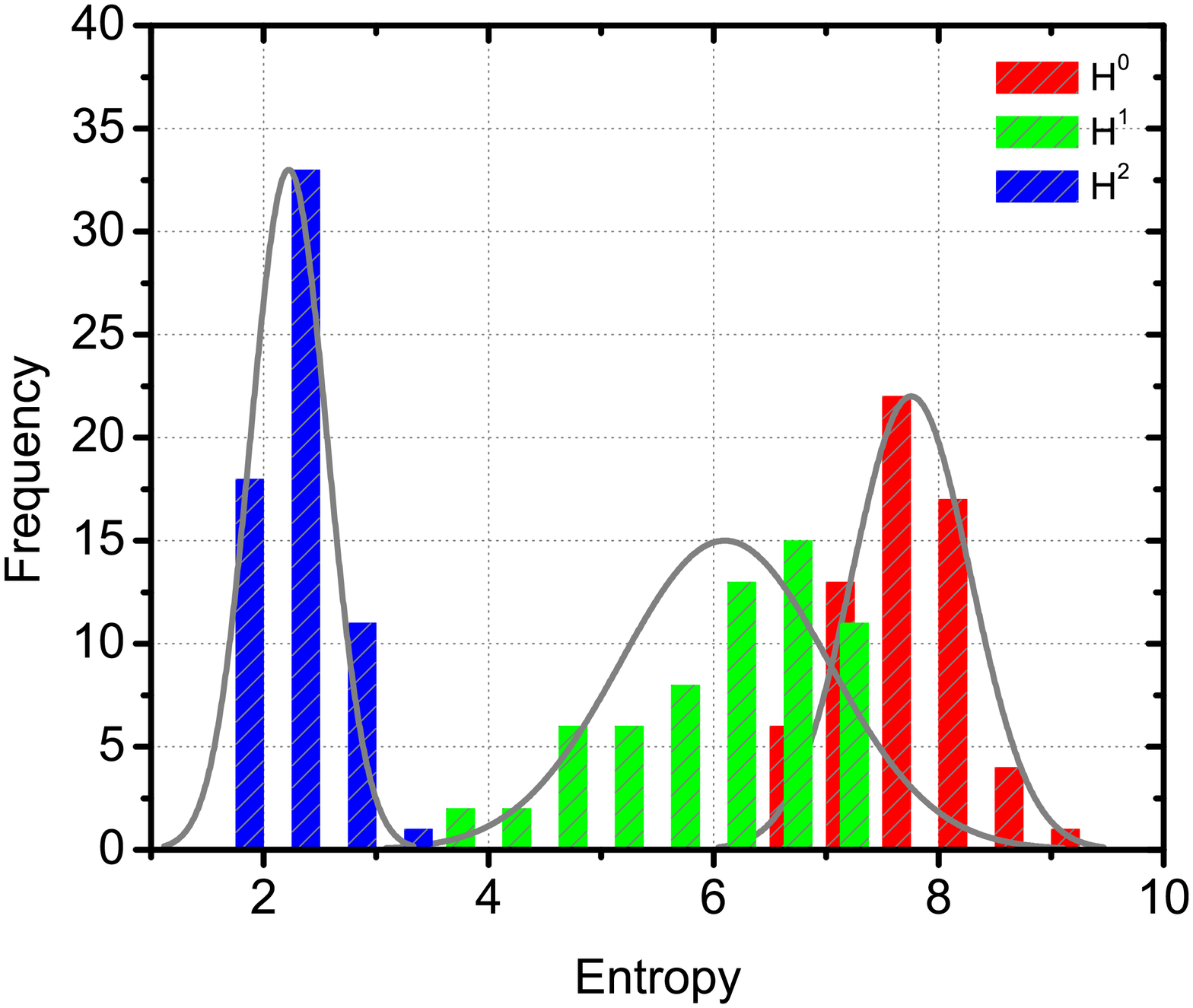}}
 \caption{Frequency count of estimated $H^0$, $H^1$ and $H^2$ from users in (a) online conversation partner sequence and (b) online location check-in place sequence.}
 \label{fig:indv_entropy}
\end{figure*}

The calculations of these quantities require an accurate estimation of the probabilities $P_A(j)$ and $P(j|i)$. However, in the absence of unlimited data, estimating these probabilities with finite sampling renders a biased estimation of the entropy, since the finite sampling makes the user activity less variable than it is, resulting in a downward bias of the entropy, and a upward bias of the the mutual information~\cite{biascorrS07}.The problems associated with estimating entropies for sparse data have been extensively explored in the literature and a variety of remedies proposed~\cite{biascorrP96}. The most common solution is to restrict the measurements to situations where one has an adequate amount of user activity data~\cite{infotransferG12}. In what follows we filter out users who are below a certain activity level,  $1000$ in our observational period. Since both $H^1$ and $H^2$ generally decrease by different amounts when taking into account finite size effects, we also performed a through bootstrap test to confirm that the empirical values of mutual information are significantly different from zero. Another widely accepted method is to estimate the magnitude of the systematic bias that originates from finite size effects and then subtract this bias from the estimated entropy. To do so, we used the Panzeri-Treves bias correction method~\cite{biascorrP96} in our calculations. The lead terms in the bias are, respectively
\begin{equation}
BIAS[{H_A}(i)] =  - \frac{1}{{2N\ln (2)}}[\overline M  - 1],
\end{equation}
and
\begin{equation}
BIAS[{H_A}(i|j)] =  - \frac{1}{{2N\ln (2)}}\sum\limits_j {[\overline {{M_j}}   - 1]},
\end{equation}
where $\overline M$ denotes the estimated number of outcome states, $\overline{M_j}$ denotes the number of different states $i$ with nonzero probability of being observed given that the previous state is $j$, and $N$ is the total number of observations. Thus, the leading term of the mutual information bias equals
\begin{equation}
BIAS[I(i;j)] = \frac{1}{{2N\ln (2)}}\{\sum\limits_j {[\overline {{M_j}}   - 1] - [\overline M  - 1]}\}.
\end{equation}
In what follows, we include the above adjustments to eliminate the impact of the finite size amount of data.

\section*{Results}

\begin{figure*}[htl]
 \centering
  \subfigure[Conversation] {\includegraphics[width=6cm]{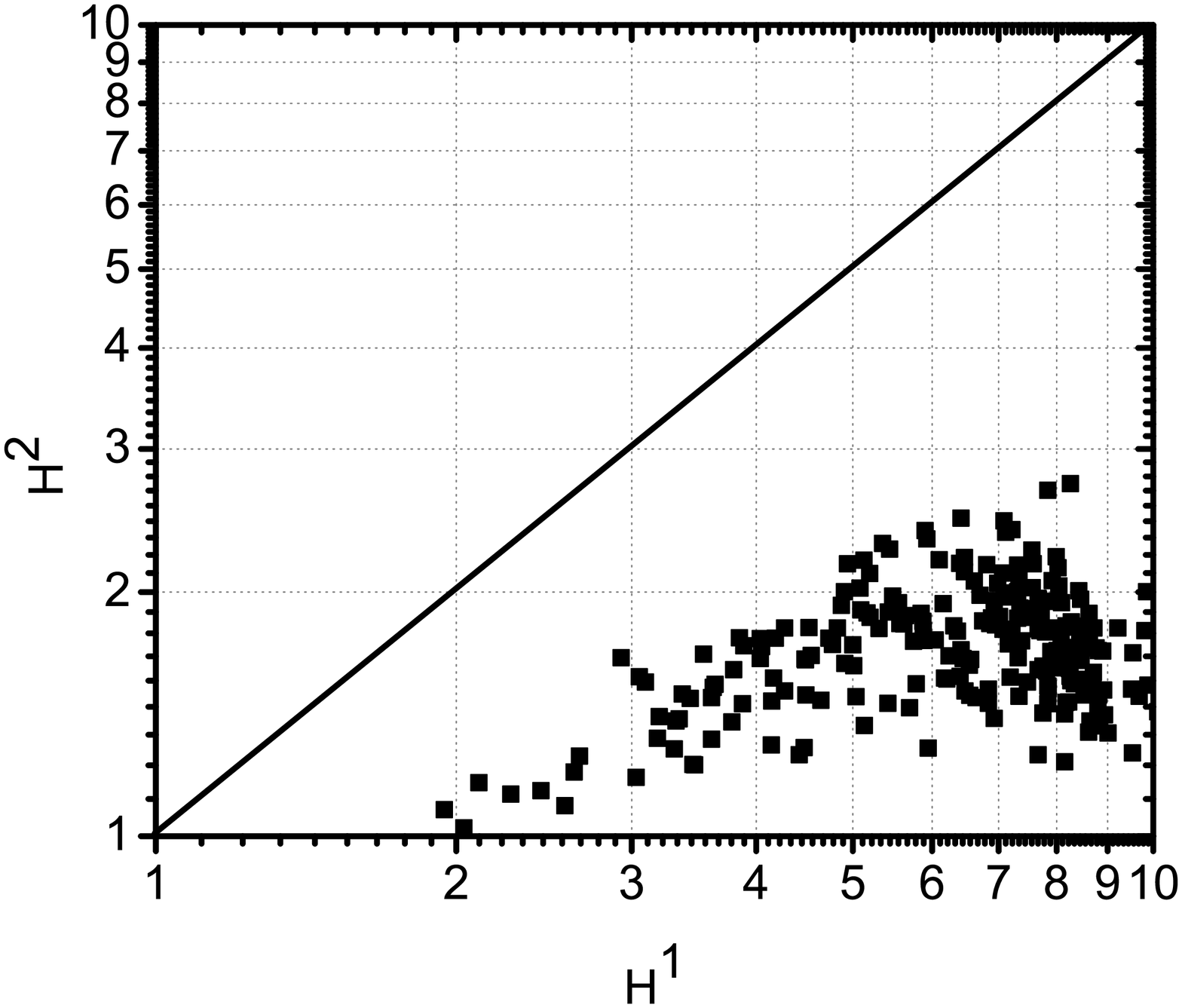}}
  \subfigure[Location Check-in]{\includegraphics[width=6cm]{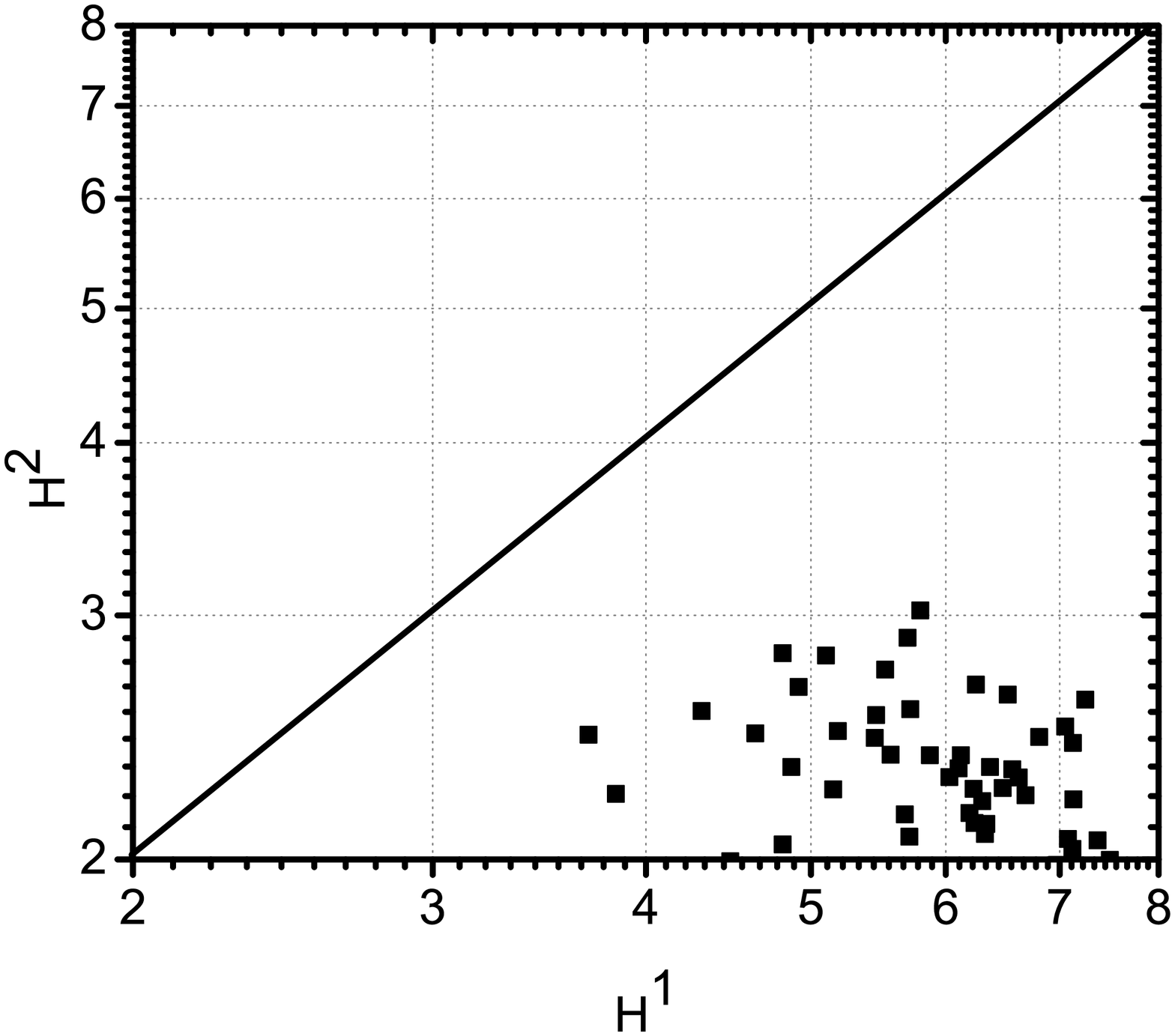}}
 \caption{Relationship between the measured $H^1$ and $H^2$ in (a) online conversations and (b) location check-ins. The solid line in the plot represents the line where $H^1$ and $H^2$ are equal. Black dots in the plot correspond to individual activity sequences.}
 \label{fig:H1_H2}
\end{figure*}

\begin{figure*}[htl]
 \centering
  \subfigure[Conversation] {\includegraphics[width=6cm]{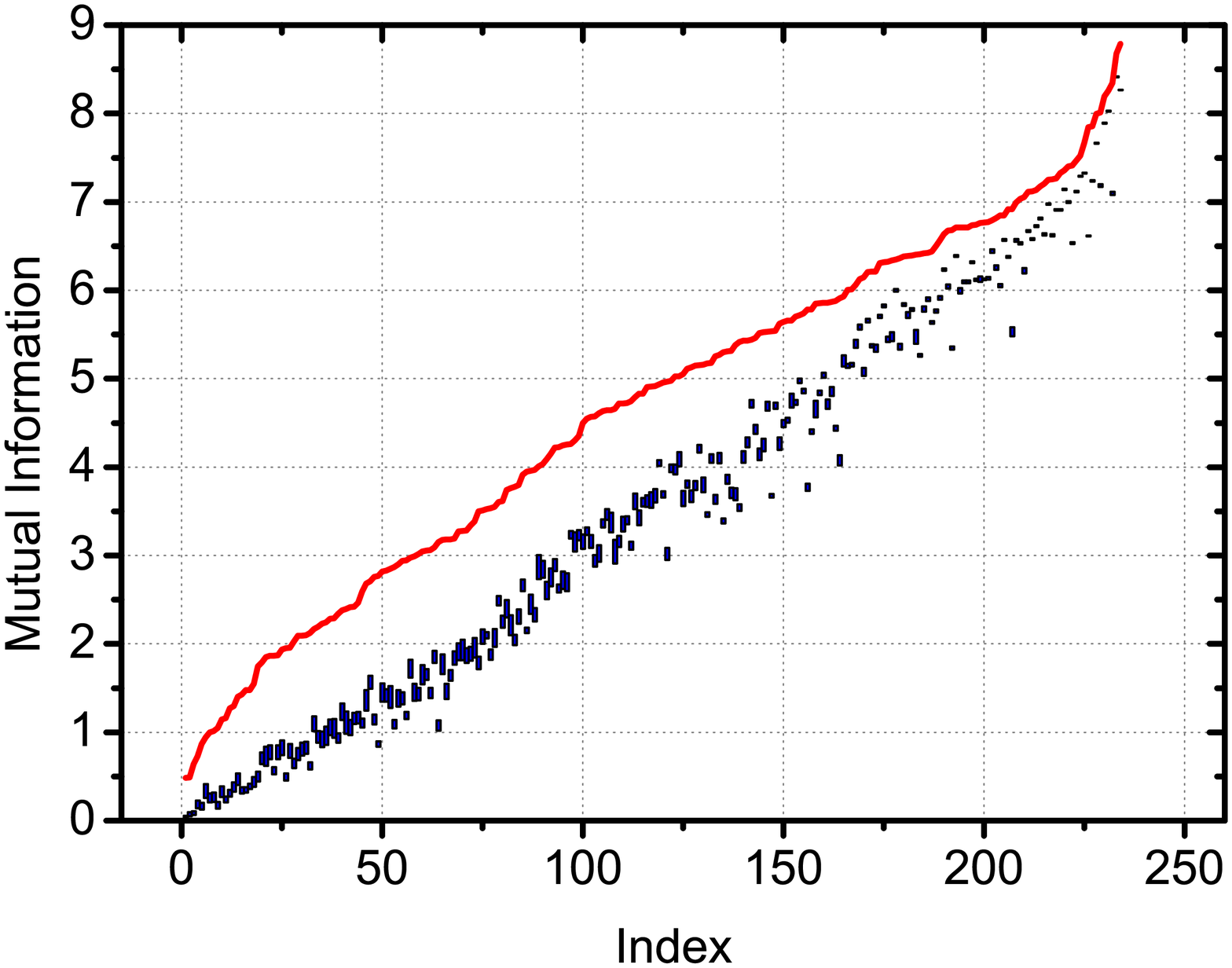}}
  \subfigure[Location Check-in]{\includegraphics[width=6cm]{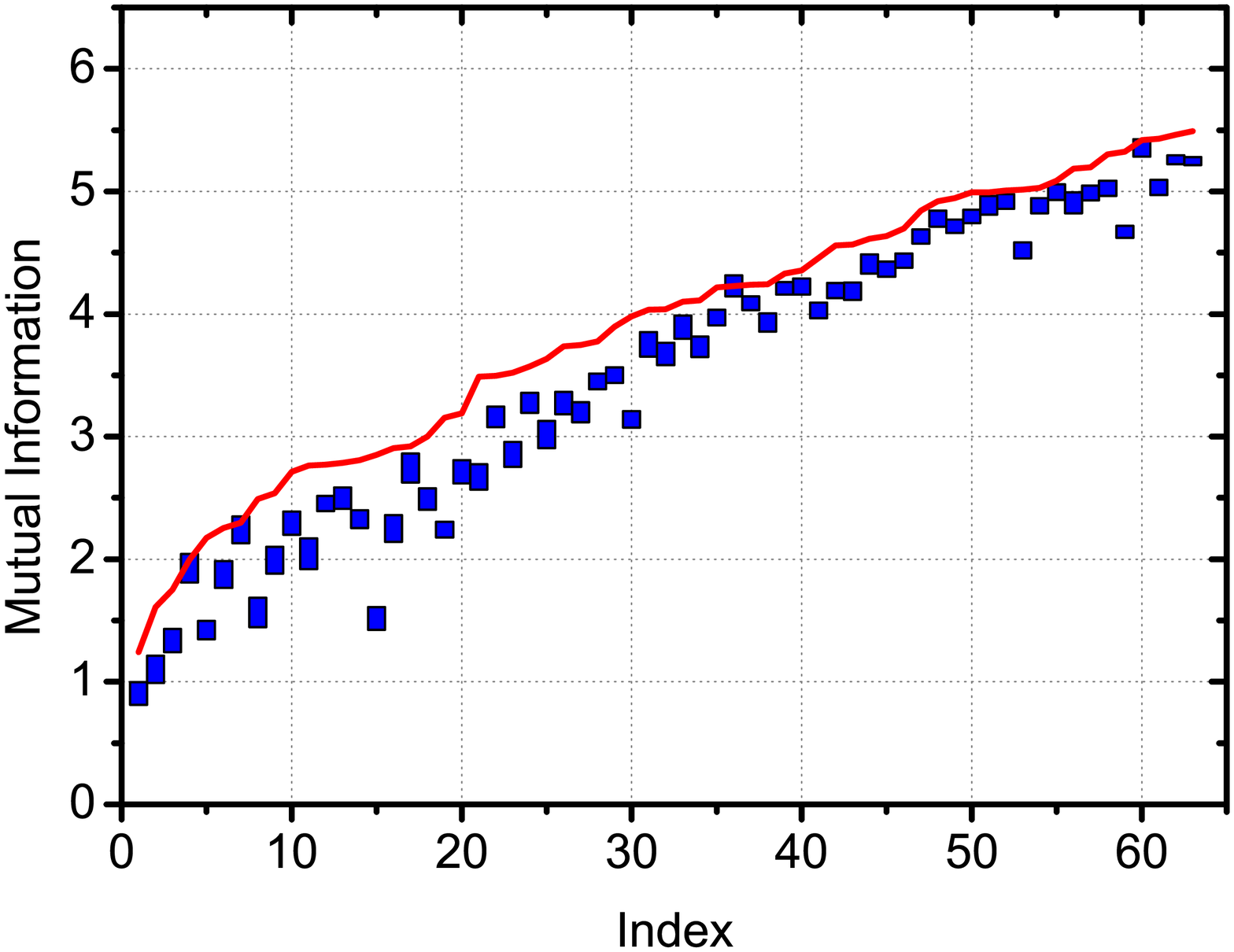}}
 \caption{Estimated mutual information and statistics of bootstrap samples. The red line is the mutual information estimated from observed online activity sequences. The upper and lower end of the blue columns represent the $2.5\%$ and $97.5\%$ percentile of $1000$ shuffled sequences for (a) online conversation partner sequences and (b) online location check-in place sequences.}
 \label{fig:mutual_info}
\end{figure*}

We start by looking at the predictability of individual activities as measured by both the entropy and the mutual information extracted from sequences in the Whrrl and Epinion datasets, respectively. The histograms of $H^0$, $H^1$ and $H^2$ calculated from user activity sequences are shown in Figure~\ref{fig:indv_entropy}. The gray solid line in the plot shows a normal fit to the frequency count. The gap between $H^1$ and $H^0$ suggests a preference for certain activities, while the difference between the values of $H^1$ and $H^2$ in the figure indicates the existence of second order correlations between states. Values of $H^1$ and $H^2$ for each individual in the online conversation network and the location check-in one are  shown in Figure~\ref{fig:H1_H2}. The straight line corresponds to  $H^1$ equal to $H^2$. One interesting fact is that all the dots are below the straight line, which confirms that there is a positive difference between $H^1$ and $H^2$ for all individuals. This difference, which is the mutual information conditioned on previous states of user activity, is plotted in increasing order as the red line in Figure~\ref{fig:mutual_info} for (a) conversations and (b) location check-ins. The positive values of the mutual information indicate information gain, or predictability, conditioned on historical states.

\begin{figure*}[htl]
 \centering
 \includegraphics[width=8cm]{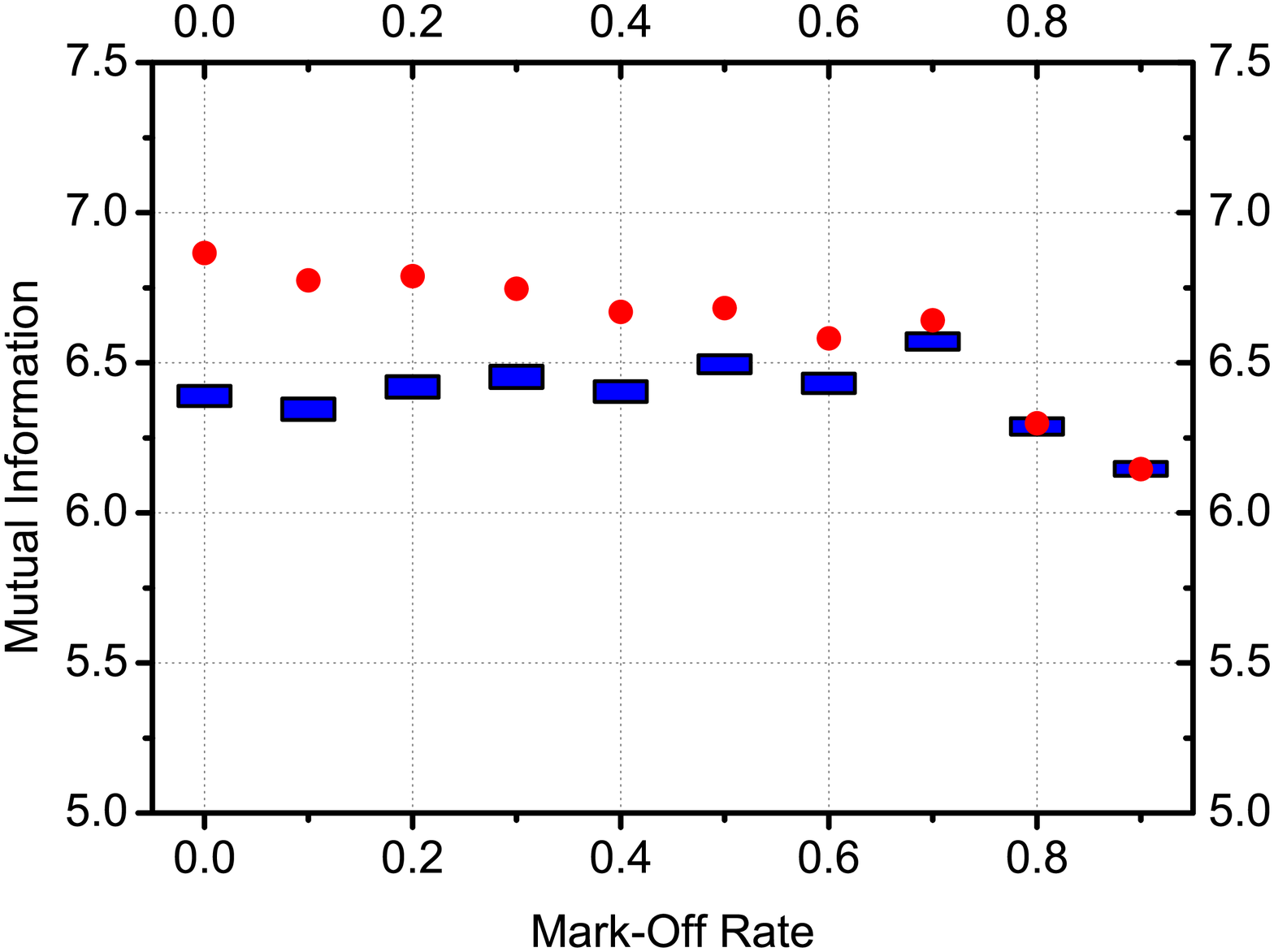}
 \caption{Mutual information and statistics of bootstrap as a function of mark-off rate. Red dots in the plot shows mutual information of sequence after mark-off. Blue bar in the plot shows the mutual information of that marked-off sequence with random shuffling. Up to $60\%$ of hidden data from the true sequence, there is deterministic pattern in the sequence after mark-off.}
 \label{fig:whrrl_markoff}
\end{figure*}

\begin{figure*}[htl]
 \centering
 \includegraphics[width=7.5cm]{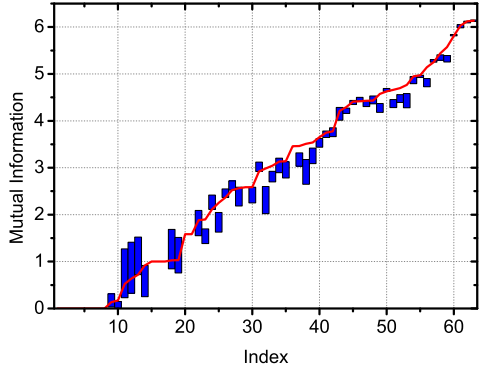}
 \caption{Estimated mutual information and statistics of bootstrap samples for group activities from Whrrl dataset. The red line is the mutual information estimated from observed online activity sequences. The upper and lower end of the blue column represent the $2.5\%$ and $97.5\%$ percentile of the shuffled sequences.}
 \label{fig:Whrrl_group_mutual_info}
\end{figure*}

\begin{figure*}[htl]
 \centering
 \includegraphics[width=7cm]{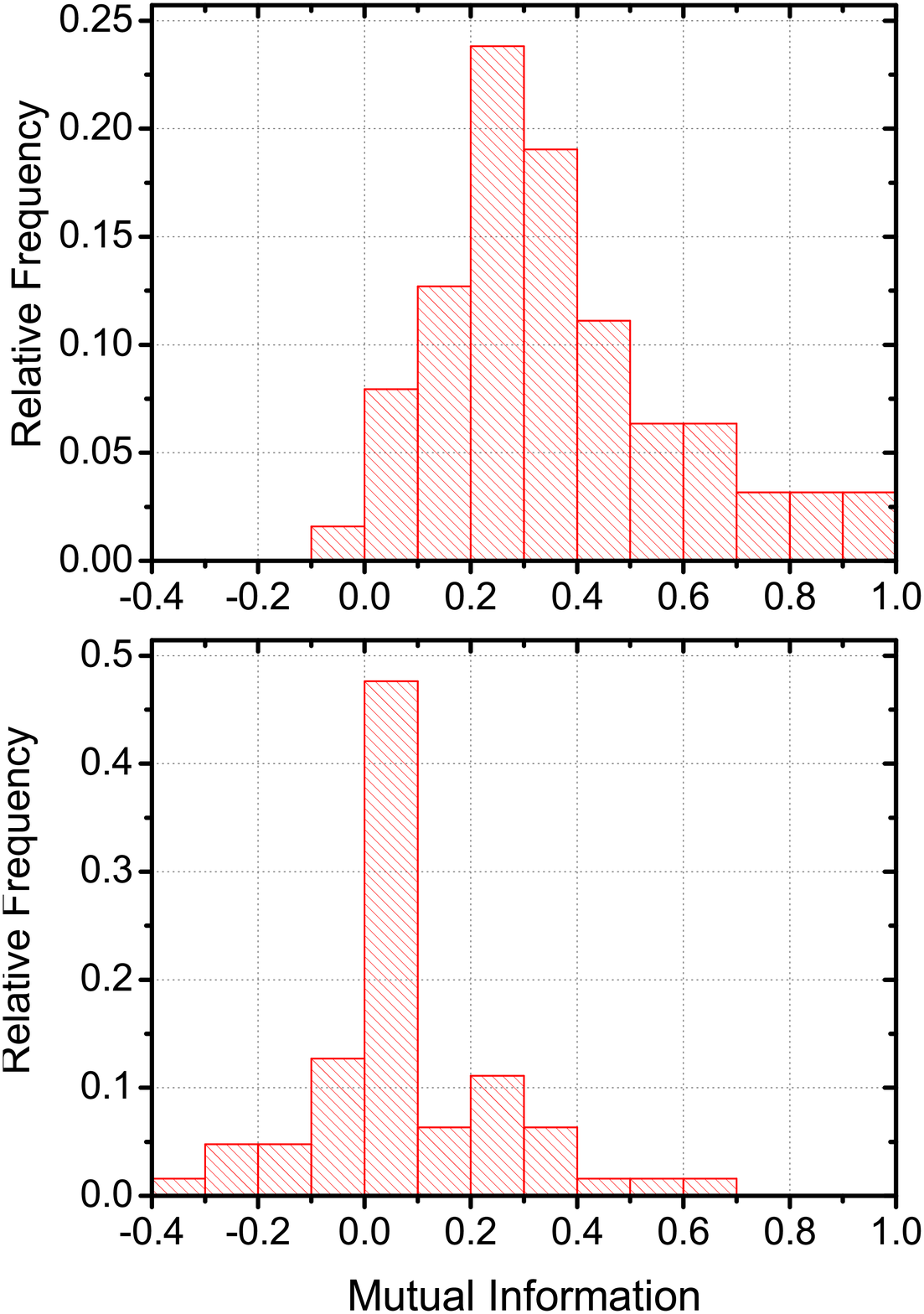}
 \caption{The upper plot shows the density of $G_{Individual}$. The lower plot shows density of $G_{Group}$. The gap for individual activities has a larger mean compared with that of attending group activities.}
 \label{fig:whrrl_indv_group_unbias}
\end{figure*}

We now examine the validity of the positive mutual information values in greater detail. There are usually two limitations when performing mutual information  measurements. The first one is the potential bias resulting from the finite data size. The second one is the possibility of missing data points in the observation process. To make sure that our results are significant and are not impacted by these two limitations, we performed the following analysis. To establish that the observed positive value of the mutual information is not due to the finite size of our data sets, we performed a bootstrap test similar to that used in human conversation studies~\cite{convpredT11}. The null hypothesis of this test is that the mutual information has a positive value because of the finite size of the dataset. For this test we set the significance level to $5\%$. We first shuffled the true activity sequence and constructed a new sequence by drawing elements randomly one by one from the original sequence without replacement. If there is a second order correlation in the original sequence the shuffled sequence breaks the order and will thus have a higher ${H_A}^2$ value, while the value of ${H_A}^1$  would be the same before and after the bootstrap. This would result in a mutual information $I_A$ value smaller than that of the true sequence. The test checks if the value of $I_A$ obtained from the original activity sequence is significantly different from the shuffled one. To obtain an estimate of the distribution for the shuffled sequence we performed the shuffling procedure a $1000$ times for each user and calculated each individual's shuffled mutual information. The value of the simulated sequence ranging from $2.5\%$ to $97.5\%$ is shown by the blue column in Figure~\ref{fig:mutual_info}. As can be seen, the red line (mutual information for true activity sequence) lies well above the upper end of the $97.5\%$ error bar, which suggests that the value of the original sequence is significantly different from that generated by the simulated sequences. We can then reject the null hypothesis at the $5\%$ significance level and conclude that the positive mutual information we obtained is not due to the limited size of the data. Furthermore, the fact that the mutual information is significantly different from zero suggests that a user's current online activity predicts his next interactions. Next, we assessed the impact of potential loss of data points in the observation period by marking off a percentage of data points from the observed location check-in sequence from Whrrl dataset. In real applications of predicting user behavior, a key question to apply maximum likelihood estimation depends on the size of observations and the ratio of missing points. To examine the impact of ratio, we perform a mark-off  on the bootstrap test of mutual information. We hide data points randomly from the true sequence while keeping the chronological order in the remaining sequence.  For example if we have a mark-off rate of 0.5, then $50\%$ of the states from the true sequence is marked off. The result of the bootstrap test after mark-off is shown in Figure~\ref{fig:whrrl_markoff}. In the plot, the red dot shows the mutual information of the true sequence after performing mark-off procedure. The thick blue bar in the plot demonstrates the mutual information of the exact same sequence with shuffling. The mutual information is significantly different from that of the random shuffled sequence until the mark-off rate reaches $60\%$. For values of the mark-off rate  larger than $60\%$, the difference between the two is broken when we fail to reject the null hypothesis that the sequence is significantly different from randomly shuffled. It is thus a confirmation that the deterministic pattern we observed is a robust one. This test also suggests the existence of a higher order correlations, larger than two, in human social online behavior. Thus, the deterministic pattern discussed in this study is a robust phenomenon which can be applied to the general situations with missing or incomplete observations. 

As mentioned earlier, we also explored whether individuals acting alone are less predictable than when becoming  members of a group. Specifically, we investigated how predictable each user's is when engaged in group activities as compared with the predictability of individual ones. In the Whrrl dataset users can expose their position with a group of other users thus providing a sequence of group attendances by users and filtering out the places that were checked in by the user alone. We then calculated the information entropies and performed the same bootstrap test as before. The calculated mutual information of the activity sequences and shuffled sequences are shown in Figure~\ref{fig:Whrrl_group_mutual_info}. Interestingly, the gap between the red line of true observation and the upper end of the error bar is is smaller than the one we obtained for the individual activities. In contrast with Figure~\ref{fig:mutual_info}(b),  the differences between the randomly shuffled sequences and the true observations are smaller. To quantify the  observed difference, we calculated the gap between the mutual information from the true activity sequence and the $97.5\%$ percentile value of the shuffled sequences, defined by $G_{Individual} = I_{Individual}-{I_{Individual}}^{0.975}$ and $G_{Group} = I_{Group}-{I_{Group}}^{0.975}$. This allows for a comparison of sequences with different lengths. The relative frequency plot of this $G_{Individual}$ and $G_{Group}$ is plotted in Figure~\ref{fig:whrrl_indv_group_unbias}. The upper plot in Figure~\ref{fig:whrrl_indv_group_unbias} shows the density plot of the gap for individual activity sequences while the lower plot shows the gap for group activities. As can be seen, the gap for individual activities has a larger of the mode compared with that of the group activities. Under the assumption that both populations from $G_{Individual}$ and $G_{Group}$ are random, independent, and arising from a normally distributed population with equal variances, the two sample t-test rejects the null hypothesis of an equal mean with a p-value of $4.88\times10^{-12}$ under $5\%$ significance level. This implies that it is harder to predict the a user's group activities than his individual ones. The values of $G_{Individual}$ versus $G_{Group}$ for each individual are plotted in Figure~\ref{fig:whrrl_indv_group_unbias}. The mean of $G_{Individual}$ is larger than $G_{Group}$. One possible explanation for this observation is that when individuals attend group activities, the decision as to what to do next is not  usually made by the individual himself. Thus, the tendency to follow others in their decisions tends to break one's regular patterns. This extra randomness would result in a larger value of ${H_A}^2$ and thus become less predictable.

\section*{Discussion}
In summary, we have shown that sequences of user online activities have deterministic components that can be used for predicting future activities. Using methods from information theory, we experimentally measured how much additional information can be gained from knowledge of previous states within a users' activity sequences. While the degree of predictability varies from person to person, we also established that it is different when individuals join a group. Besides the intrinsic interest of these findings, the fact that one can predict online social interactions should be helpful in improving the design of algorithms and applications for online social sites.

\section*{Acknowledgements}
C. W. would like to thank HP Labs for financial support. 

\section*{Author contributions}
C.W. and B.H. conceived and designed the study. C.W. collected the data. C.W. and B.H. discussed the results and wrote the manuscript.

\section*{Additional information}
\subsection*{Competing financial interests}
The authors declare no competing financial interests.


\begin{thebibliography}{10}

\bibitem{webB01}
B.~A. Huberman.
\newblock The laws of the web: Patterns in the ecology of information.
\newblock {\em The MIT Press}, 2001.

\bibitem{humanV06}
Alexei V\'{a}zquez, Jo\'{a}o Gama~Oliveira, Zolt\'{a}n Dezs\'{o}, Kwang-Il Goh,
  Imre Kondor, and Albert-L\'{a}szl\'{o} Barab\'{a}si.
\newblock Modeling bursts and heavy tails in human dynamics.
\newblock {\em Physical Review E}, \textbf{73}(3):036127, 2006.

\bibitem{webB06}
Scott~A. Golder, Dennis~M. Wilkinson, and Bernardo~A. Huberman.
\newblock Rhythms of social interaction: messaging within a massive online
  network.
\newblock {\em International Conference on Communities and Technologies}, 2007.

\bibitem{webF07}
Fang Wu and B.~A. Huberman.
\newblock Novelty and collective attention.
\newblock {\em Proc. Natl. Acad. Sci.}, \textbf{105}(17599), 2007.

\bibitem{humanM08}
M.~C. Gonz\'{a}lez, C.~A. Hidalgo, and A.-L. Barab\'{a}si.
\newblock Understanding individual human mobility patterns.
\newblock {\em Nature (London)}, \textbf{453}:779, 2008.

\bibitem{webR10}
Jacob Ratkiewicz, Santo Fortunato, Alessandro Flammini, Filippo Menczer, and
  Alessandro Vespignani.
\newblock Characterizing and modeling the dynamics of online popularity.
\newblock {\em Phys. Rev. Lett.}, \textbf{105}(15):158701, 2010.

\bibitem{humanW10}
Ye~Wu, Changsong Zhou, Jinghua Xiao, J{u}rgen Kurths, and Hans~Joachim
  Schellnhuber.
\newblock Evidence for a bimodal distribution in human communication.
\newblock {\em Proc. Natl. Acad. Sci.}, \textbf{107}:18803, 2010.

\bibitem{webS11}
Scott~A. Golder and Michael~W. Macy.
\newblock Diurnal and seasonal mood vary with work, sleep, and day length across
  diverse cultures.
\newblock {\em Science}, \textbf{333}(6051), 2011.

\bibitem{webH98}
B.~A. Huberman, P.~L.~T. Pirolli, J.~E. Pitkow, and R.~M. Lukose.
\newblock Strong regularities in world wide web surfing.
\newblock {\em Science}, \textbf{280}(95), 1998.

\bibitem{webT03}
Joshua~R. Tyler and John~C. Tang.
\newblock When can i expect an email response? a study of rhythms in email
  usage.
\newblock In {\em ECSCW}, pages 239--258. Springer, 2003.

\bibitem{webG11}
Bruno Gon�alves, Nicola Perra, and Alessandro Vespignani.
\newblock Modeling users' activity on twitter networks: Validation of dunbar's
  number.
\newblock {\em PLoS ONE}, \textbf{6}(8), 2011.

\bibitem{groupW07}
Stefan Wuchty, Benjamin~F. Jones, and Brian Uzzi.
\newblock The increasing dominance of teams in production of knowledge.
\newblock {\em Science}, \textbf{316}(5827):1036--1039, 2007.

\bibitem{groupP07}
G.~Palla, A.~Barabasi, and T.~Vicsek.
\newblock Quantifying social group evolution.
\newblock {\em Nature}, \textbf{446}:664--667, 2007.

\bibitem{groupD09}
Munmun~De Choudhury.
\newblock Modeling and predicting group activity over time in on-line social
  media.
\newblock In {\em Hypertext}, pages 349--350, 2009.

\bibitem{humanS03}
S.~L. Scott and P.~Smyth.
\newblock The markov modulated poisson process and markov poisson cascade with
  applications to web traffic data.
\newblock {\em Bayesian Statistics}, \textbf{7}, 2003.

\bibitem{burstyE04}
J.P. Eckmann, E.~Moses, and D.~Sergi.
\newblock Entropy of dialogues creates coherent structures in e-mail traffic.
\newblock {\em Proc. Natl. Acad. Sci.}, \textbf{101}(14333), 2004.

\bibitem{burstyB05}
A.L. Barabasi.
\newblock The origin of bursts and heavy tails in human dynamics.
\newblock {\em Nature}, \textbf{435}, 2005.

\bibitem{burstyR09}
D.~Rybski, S.~Buldyrev, S.~Havlin, F.~Liljeros, and H.~Makse.
\newblock Scaling laws of human interaction activity.
\newblock {\em Proc. Natl. Acad. Sci.}, \textbf{106}(12640), 2009.

\bibitem{burstyC08}
R.~Crane and D.~Sornette.
\newblock Robust dynamic classes revealed by measuring the response function of
  a social system.
\newblock {\em Proc. Natl. Acad. Sci.}, \textbf{105}(15649), 2008.

\bibitem{burstyE09}
Eytan Adar, Jaime Teevan, and Susan Dumais.
\newblock Resonance on the web: Web dynamics and revisitation patterns.
\newblock In {\em CHI}, 2009.

\bibitem{burstyA10}
Anna Chmiel, Kamila Kowalska, and Janusz~A. Holyst.
\newblock Scaling of human behavior during portal browsing.
\newblock {\em Phys. Rev. E}, \textbf{80}(066122), 2010.

\bibitem{convpredT11}
Taro Takaguchi, Mitsuhiro Nakamura, Nobuo Sato, Kazuo Yano, and Naoki Masuda.
\newblock Predictability of conversation partners.
\newblock {\em Phys. Rev. X}, \textbf{1}(011008), 2011.

\bibitem{mobpredS10}
Chaoming Song, Zehui Qu, Nicholas Blumm, and Albert-L\'{a}szl\'{o}
  Barab\'{a}si.
\newblock Limits of predictability in human mobility.
\newblock {\em Science}, \textbf{327}(5968):1018--1021, 2010.

\bibitem{rwP98}
L.~Page, S.~Brin, R.~Motwani, and T.~Winograd.
\newblock The pagerank citation ranking: Bringing order to the web.
\newblock Technical report, Stanford University, 1998.

\bibitem{rwB08}
Shumeet Baluja, {\em et al}.
\newblock Video suggestion and discovery for youtube: taking random walks
  through the view graph.
\newblock In {\em WWW}, pages 895--904. ACM, 2008.

\bibitem{rwA10}
Nikolay Archak, Vahab~S. Mirrokni, and S.~Muthukrishnan.
\newblock Mining advertiser-specific user behavior using adfactors.
\newblock In {\em WWW}, pages 31--40. ACM, 2010.

\bibitem{linkL11}
Linyuan Lu and Tao Zhou.
\newblock Link prediction in complex networks: A survey.
\newblock {\em Physica A: Statistical Mechanics and its Applications},
  \textbf{390}(6):1150--1170, 2011.

\bibitem{webmarkovF12}
Flavio Chierichetti, Ravi Kumar, Prabhakar Raghavan, and Tam\'{a}s Sarl\'{o}s.
\newblock Are web users really markovian?
\newblock In {\em WWW}, pages 609--618. ACM, 2012.

\bibitem{mutualinfoS02}
Ralph~E. Steuer, J\'{u}rgen Kurths, Carsten~O. Daub, Janko Weise, and Joachim
  Selbig.
\newblock The mutual information: Detecting and evaluating dependencies between
  variables.
\newblock In {\em ECCB}, pages 231--240, 2002.

\bibitem{biascorrS07}
Marcelo A.~Montemurro Stefano~Panzeri, Riccardo~Senatore and Rasmus~S.
  Petersen.
\newblock Correcting for the sampling bias problem in spike train information
  measures.
\newblock {\em J Neuroiol}, \textbf{3}(98):1064--1072, 2007.

\bibitem{biascorrP96}
S.~Panzeri and A.~Treves.
\newblock Analytical estimates of limited sampling biases in different
  information measures.
\newblock {\em Network: Computation in Neural Systems}, \textbf{7}:87107, 1996.

\end{thebibliography}
\end{document}